# The Large Area Detector of LOFT: the Large Observatory for X-ray Timing

S. Zane*[1], D. Walton[1], T. Kennedy[1], M. Feroci[2], J.-W. Den Herder[3], M. Ahangarianabhari[4], A. Argan[5], P. Azzarello[6], G. Baldazzi[7], M. Barbera[23,24], D. Barret[8], G. Bertuccio[4], P. Bodin[9], E. Bozzo[6], L. Bradley[1], F. Cadoux[10], P. Cais[11], R. Campana[13], J. Coker[1], A. Cros[8], E. Del Monte[2], A. De Rosa[2], S. Di Cosimo[2], I. Donnarumma[2], Y. Evangelista[2], Y. Favre[10], C. Feldman[12], G. Fraser[12]•, F. Fuschino[13], M. Grassi[14], M.R. Hailey[1], R. Hudec[15], C. Labanti[13], D. Macera[4], P. Malcovati[14], M. Marisaldi[13], A. Martindale[12], T. Mineo[16], F. Muleri[2], M. Nowak[17], M. Orlandini[13], L. Pacciani[2], E. Perinati[18], V. Petracek[19], M. Pohl[10], A. Rachevski[20], P. Smith[1], A. Santangelo[18], J.-Y. Seyler[11], C. Schmid[21], P. Soffitta[2], S. Suchy[18], C. Tenzer[18], P. Uttley[22], A. Vacchi[20], G. Zampa[20], N. Zampa[20], J. Wilms[21], B. Winter[1]

on behalf of the LOFT Consortium

[1]Mullard Space Science Laboratory, UCL, Holmbury St Mary, Dorking, Surrey, RH56NT,UK, [2]INAF-IAPS-Roma via Fosso del Cavaliere, 100, 00133, Rome, Italy, [3]SRON, The Netherlands Institute of Space Research, Utrecht, The Netherlands, [4]Politecnico di Milano, Como campus, via Anzani 42, 22100, Como, Italy, [5]INAF HQ, Viale del Parco Mellini 84, 00136, Rome, Italy, [6]ISDC, Science Data Center for Astrophysics, Ch. D'Ecogia 16, 1290, Versoix, Geneva, Switzerland, [7]University of Bologna, Dept. of Physics and INFN section of Bologna, V.le Berti Pichat, 6/2, 40127, Bologna, Italy, [8]Institut de Recherche en Astrophysique et Planetologie, IRAP, 9 Avenue du Colonel Roche, BP44346, 31028, Toulouse, France, [9]Centre National d'Etudes Spatiales, Centre Spatial de Toulouse, 18 Avenue Edouard Belin, 31 401, Toulouse, CEDEX 9, France, [10]DPNC, Geneva University, Quai Ernest-Ansermet 24, CH-1211,Geneva, Switzerland, [11]Laboratoire d'Astrophysique de Bordeaux, Univ. Bordeaux, CNRS, UMR5804, BP 89, 33270 Floirac – France, [12]Space Research Centre, Department of Physics and Astronomy, University of Leicester, Leicester, LE17RH, UK, [13]INAF/IASF Bologna, via Gobetti 101, 40129, Bologna, Italy, [14]University of Pavia, via Ferrata 1, 27100, Pavia, Italy, [15]Czech Technical University in Prague, Faculty of electrical Engineering and Astronomical Institute, Academy of Science of the Czech Republic, Ondrejov, Czech Republic, [16]INAF/IASF Palermo, via Ugo la Malfa 153, 90146, Palermo, Italy, [17]MIT, NE80-6077, 77 Massachusetts Ave., Cambridge, MA 02139, US, [18]IAAT, University of Tuebingen, Sand 1, 72076, Tuebingen, Germany, [19]Czech Technical University in Prague, Faculty of Nuclear Science, Prague, Czech Republic, [20]Istituto Nazionale di Fisica Nucleare, INFN, Sezione di Trieste, Padriciano 99, I-34149, Trieste, Italy, [21]Dr Remeis-Observatory & ECAP, University of Erlangen-Nuremberg, Sternwartstr. 7, 96049 Bamberg, Germany, [22]Astronomical Institute Anton Pannokoek, University of Amsterdam, Postbus 94249, 1090 GE Amsterdam, The Netherlands, [23]Dipartimento di Fisica e Chimica, Palermo University, Via Archirafi, 36 - 90123 Palermo, Italy, [24]INAF-Osservatorio Astronomico di Palermo, Piazza del Parlamento 1, 90134 Palermo, Italy.

*s.zane@ucl.ac.uk; phone 0044 1483204282; fax 0044 1483 278312

•This work is dedicated to the memory of Prof. George Fraser, a dear friend and esteemed colleague.

**ABSTRACT**

LOFT (Large Observatory for X-ray Timing) is one of the five candidates that were considered by ESA as an M3 mission (with launch in 2022-2024) and has been studied during an extensive assessment phase. It is specifically designed to perform fast X-ray timing and probe the status of the matter near black holes and neutron stars. Its pointed instrument is the Large Area Detector (LAD), a 10 $m^2$-class instrument operating in the 2-30keV range, which holds the capability to revolutionise studies of variability from X-ray sources on the millisecond time scales.

The LAD instrument has now completed the assessment phase but was not down-selected for launch. However, during the assessment, most of the trade-offs have been closed leading to a robust and well documented design that will be re-proposed in future ESA calls. In this talk, we will summarize the characteristics of the LAD design and give an overview of the expectations for the instrument capabilities.

**Keywords:** X-ray, Timing, Compact Objects

## 1. INTRODUCTION

High time resolution X-ray observations of compact objects provide direct access to strong field gravity, black hole masses and spins, and the equation of state of ultradense matter, hence inputs to particle physics not testable in the laboratory and unique tests of general relativity.

LOFT (the Large Observatory for X-ray Timing) was one of the candidates originally considered by ESA as an M3 mission [1,7]. The LOFT payload consists of two instruments (Fig. 1). The Large Area Detector (LAD) is the largest instrument with an effective area reaching 10 $m^2$ (on the 6 deployable panels in the baseline consortium configuration), i.e. 20 times larger than the area of the best past timing mission (such as RXTE, the largest predecessor). The Wide Field Monitor observes a large fraction of the sky for changes in the state of targets of interest (especially NS and BHs, but AGNs as well) or to discover and localize new sources. If a relevant transient is detected, the viewing direction of LOFT can be modified to observe this source with the LAD. The sky region imaged by the WFM includes the LAD FoV, enabling arc-min imaging of its field of view (to identify possible contaminating sources). In this paper we will focus on the design description of the LAD. Thanks to its unprecedentedly large collecting area, the LAD holds the capability to revolutionize the studies of X-ray variability from X-ray sources on the millisecond time scales. The LAD is designed to operate in the energy range 2-30 keV (up to 80 keV in expanded mode) with good spectral resolution (<240 eV @ 6 keV Full Width Half Maximum, FWHM) and a temporal resolution of 10μs.

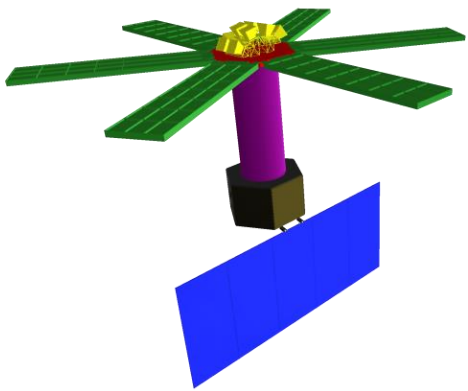

Figure 1. The configuration for LOFT as in the M3 proposal. Green = LAD, yellow = WFM, Red = Optical bench, Purple = Structural Tower, Gold = Bus, Blue = Solar array.

The key to the 20x breakthrough in effective area that can be achieved by the LAD resides in the synergy between technologies imported from other fields of scientific research, both ground- and space-based. In particular, the crucial ingredients for a sensitive but lightweight experiment, enabling ~18 $m^2$ geometric area payload at reasonable weight, are the innovative large-area Silicon Drift Detectors (SDDs) designed on the heritage of the ALICE experiment at

CERN/LHC. These are combined with a collimator based on lead-glass micro-capillary plates (the mechanical structure of the well-known microchannel plates).

The drift concept makes the spectroscopic performance of the SDDs weakly dependent on the extent of the collecting surface: large-area (~76 cm$^2$) monolithic detectors can be designed, with only 224 read-out anodes (thus low power requirements, ~20 Wm$^{-2}$) but still very good spectral performance. This design allows to achieve, for the first time, an unprecedented large throughput (~2.5 x 10$^5$ s$^{-1}$ from the Crab) with a segmented detector, making pile-up and dead-time, often worrying or limiting focused experiments, secondary issues.

The study of the energy-resolved timing properties of the X-ray emission of cosmic sources requires the accurate measurement of the time-of-arrival (TOA) and energy of the largest number of photons from the target source. The unambiguous identification of the target source in this type of experiment (e.g., the PCA on-board RXTE) is most effectively achieved by narrowing the field of view by means of an aperture collimator, down to a level (typically <~1°) large enough to allow for pointing uncertainties yet small enough to reduce the aperture background (cosmic diffuse X-ray background) and the risk of source confusion.

In this type of instrument, the knowledge of the impact point of the photon on the detector array is not needed, so there is no need for position sensitive detectors. Instead, detector read-out segmentation is useful/necessary to reduce the effects of pile-up and dead time.

The LAD is therefore designed as a classical collimated experiment.

The requirement to acquire X-ray events with high time and spectral resolution and high statistical significance (to study astrophysical sources in the time/frequency/energy domain) has led to the LAD instrument described here. The primary features are:

- A large effective area for a single on-axis source;
- No imaging, but instead collimation to maximise on-axis effective area while rejecting off-axis flux (other sources, background);
- High energy resolution.

These requirements are most efficiently realised in the LOFT design.

Collimation is supplied in a mass-, volume- and cost- effective way by micro-pore optics (MPO). Similar considerations, together with the energy resolution requirement lead to the choice of Silicon Drift Detectors (SDD) behind the MPOs. The effective area requirement entails a geometric area of ~18 m$^2$ or an active area of ~15 m$^2$.

The large effective area requirement lead to the design of the LAD payload as a large array of X-ray detectors. The current baseline for the configuration, assumed by the consortium, is based on 6 detector panels (approximately 1x3 m$^2$ each), connected by hinges to an optical bench at the top of a tower (see Figure 1 and Figure 2). The basic LAD detection element is composed of SDD + Front-End Electronics (FEE) + Collimator. The 6 Detector Panels (DP) will be tiled with 2016 detectors, electrically and mechanically organized in groups of 16, referred to as Modules (Figure 2 and Figure 3). Each of the 6 Panels hosts 21 Modules, each one in turn composed of 16 Detectors. The assembly philosophy employs a hierarchical approach: Detector, Module, Detector Panel and LAD Assembly. The read-out electronics are organized as follows: the FEEs of the 16 Detectors in a Module converge into a single Module Back End Electronics (MBEE, made up of two sections). One Panel Back-End Electronics (PBEE) for each DP is in charge of interfacing in parallel the 21 MBEE included in a PBEE, making the Module the basic redundant unit.

The instrument design is described in more detail in the next sections and in [6]. The performance requirements which LAD will meet in order to address the scientific objectives are described in [7].

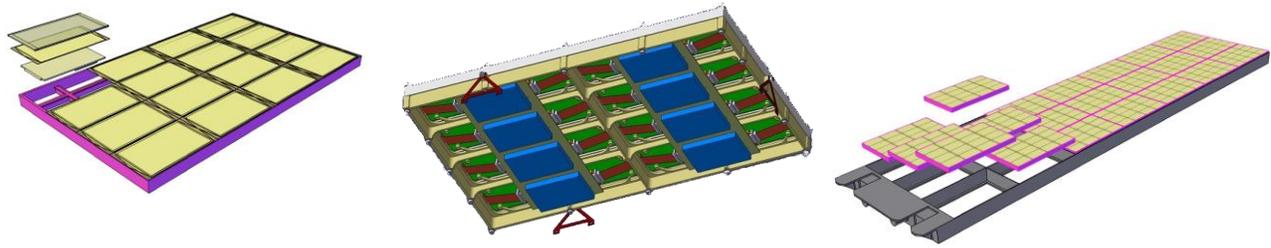

Figure 2. Left: Front-side view of a Module, showing the mounting of the collimator, SDD and the Front End Electronics. Center: Back-side view of a Module, showing the Module Back End Electronic. Right: a LOFT Detector Panel with all the assembled Modules and the interfaces to the deployment system.

## 2. LAD MECHANICAL AND THERMAL DESIGN

### 2.1 Module Mechanical Design

As mentioned in the previous section, the basic LAD detection element is the Detector, composed of SDD+FEE+Collimator. The LOFT detector module assembly comprises sixteen SDD's arranged in a four by four pattern, see Figure 3. Each of these detectors has a single SDD tile mounted on top of the FEE board with wire bonded electrical connections running around the edges. Flexi circuit ribbons run from one end of the FEE board to Hyperstack connectors on the back end electronics (BEE) and HV boards.

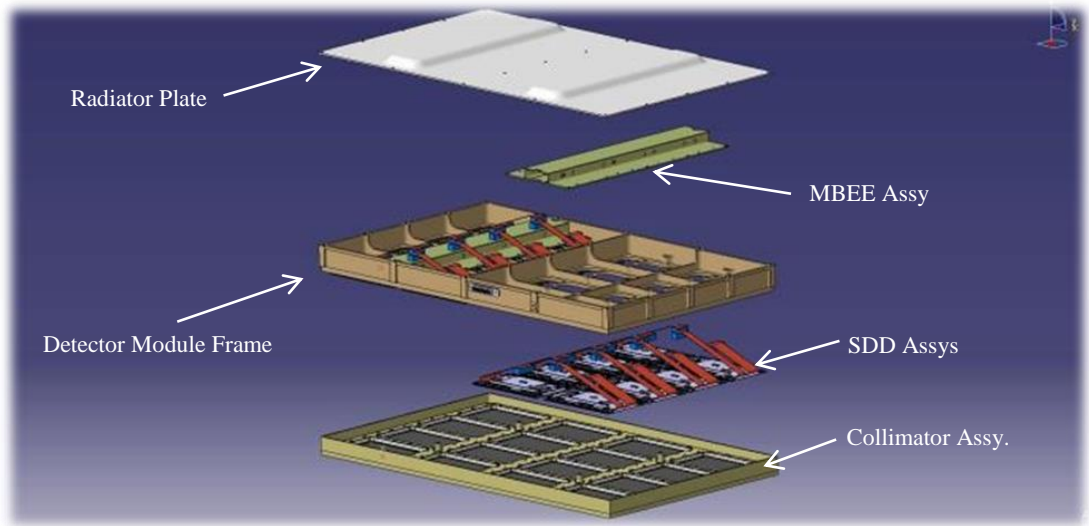

Figure 3: Exploded View of Lower Detector Module.

The collimator over each individual detector is a single micro-pore plate (MPP) 110.0 x 71.5 x 5 mm that fits into a recess in the aluminum alloy collimator frame. Due to the difficulty to achieve adequate stiffness in the collimator frame in the weight and space available, a new brazed frame design has being investigated. This frame features hollow tubular box section members to give maximum stiffness for minimum weight and can be fabricated from machined channel section halves that are then dip brazed together and finish machined. A system of pressure plates and springs has been designed in order to help fix the collimator in place: this guarantees sufficient preload, adequately spread on the MPO tile to prevent it from moving out of plane during vibration testing (see Figure 4 and details in [6]). This system also makes for faster assembly and leaves the front collimator frame face flush while permitting sufficient lateral movement to handle thermal expansions (CTE) differences. This clamping system will be further optimized, and, as part of the assessment phase activities, two brazed collimator frames have been manufactured to confirm FE analysis, practicality of manufacture and design of the spring system (see Figure 5).

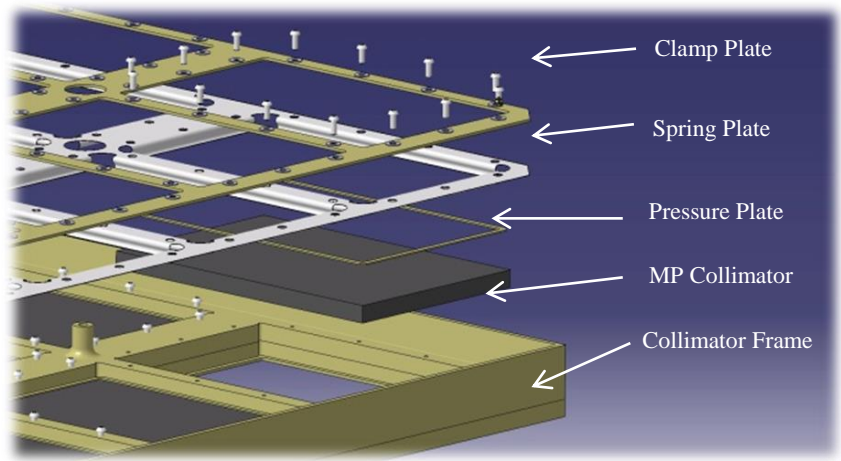

Figure 4: Exploded view on back of Collimator.

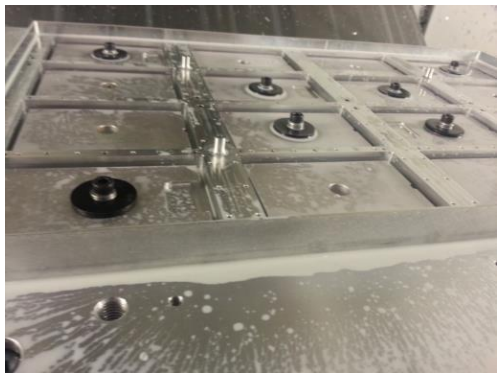

Figure 5: Collimator frame prototype.

At present the baseline is to have an optical filter mounted between the detector and the micro pore plate. Due to envelope constraints this filter will be mounted on the collimator frame, above each detector. Design of these frames needs to be progressed paying attention to purging the micro pore tiles as well as the acoustic loading of these particularly thin filters (a test assembly has been produced for early acoustic testing and the brazed frame structures are well advanced and have been successfully used for vibration testing and proof of analytical results).

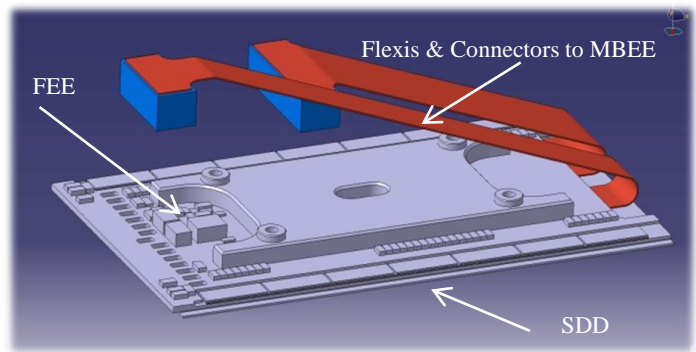

Figure 6: Silicon Drift Detector

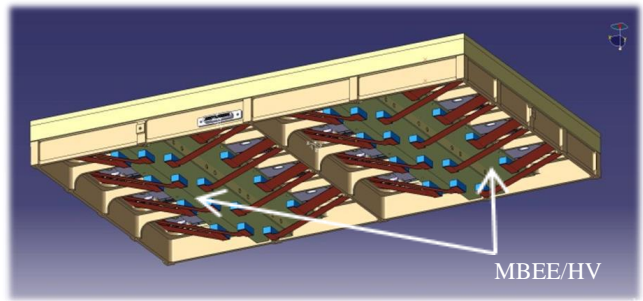

Figure 7: Bottom view of the module with the split MBEE.

The plan is then to place the whole assembly collimator face down onto a purpose made surface plate, following a concept that has been conceived to try to prevent any "locked in torque" in the final assembly that could induce distortion of the module and compromise the co-alignment between the different MP collimating tiles.

The detectors are mounted in a monolithic aluminum alloy frame that supports each detector unit on four points with sufficient compliance to cater for CTE variation between the frame and the detector FEE PCB. For each detector there is a conductive link via the structure to the radiator. The SDD detector assembly itself consists of the SDD, FEE PCB and the flying leads that connect with either one of the MBEE boards and HV units. This assembly is oriented such that the integration of each module is kept as modular as possible, without creating a need to introduce different interfaces between MBEE and the detector assembly related to their relative position inside the module. For this reason the MBEE is split into two PCBs, see Figure 7. Each of the two sub-modules consists of a group of 8 detectors with all related front end electronics. The harnesses are then combined into a single interface on the outside of the module box.

By maintaining in effect this internal modular approach it will be relatively straight forward to introduce other configurations with either 3 or 4 of these sub-modules. The possible permutations have to be limited, since they will inevitably increase overall cost, but this solution keeps open the possibility for further panel optimisation according to the design for the panels produced by a potential prime contractor (which may differ from the consortium baseline).

Two back end electronics (MBEE) PCB assemblies with HV supplies attached are then mounted behind the bank of detectors, each BEE/HV assembly covering eight detectors. A 2 mm thick aluminum alloy radiator panel is screwed directly to the back of the detector module body completing the build of the lower detector module assembly. The radiator thickness is selected to minimise the gradient across this radiator and therefore optimises the detector temperature over each orbit.

The overall mechanical design of the FEE assembly (SDD+PCB+Carbon Fiber Reinforced Plastic back support) has also been studied in detail during the assessment phase. In this paper we do not enter into details, and we just show the latest CAD model with some of the relevant details on the PCB shape and flexes (Figure 8). Particular attention has been payed to: i) the mechanical alignment of the ASIC pads, ii) thermo-mechanical stresses that can arise due to the combination of very low survival temperatures (down to -70°C, on orbit) and the stacking of different materials in the FEE assembly, and iii) mechanical stresses due to the launcher.

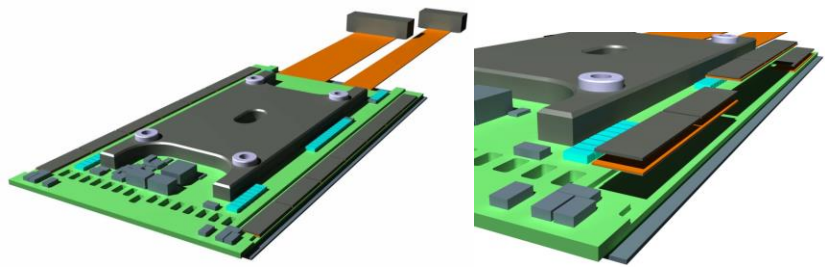

Figure 8 Left: Exploded CAD view (details on ASIC's Hybrid). Right: FEE assembly CAD views on Catia V5R20.

The thermal model of the LAD module has also evolved over the past two years addressing changes in orbit and thermal optical base line. Currently the module consists of a box holding the detectors and electronics, with a frame in front of it, holding the micro –pore optics and a radiator at the back. The box is effectively covered in second surface mirror tape, where possible. This means the radiator is covered as well as the frame holding the MPO tiles. This was introduced to optimise heat rejection combined with minimising solar heat absorption. Furthermore, the thermal link between each detector and the radiator has been maximised. An extensive design trade off was performed to assess optimal configurations for the module from a thermal point of view: a range of steady state analyses have been performed using a detailed thermal model, aimed at sizing gradients across the detectors as well as the structure, MPO and radiator.

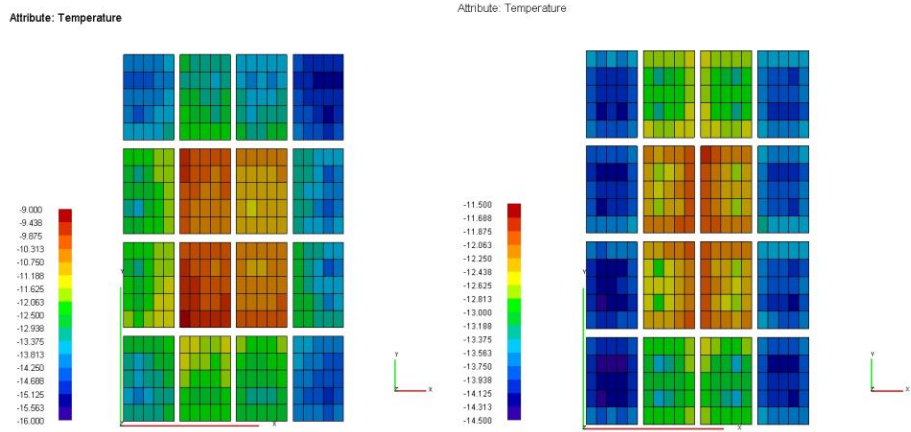

Figure 9: Detector average temperatures at extreme pointing.

As an example, Figure 9 shows on the left the detector average temperatures over one orbit with the MPO facing the Sun at a 45-deg angle. On the right the detector average temperature is shown with the radiator pointing at a 75-deg angle with respect to the orbital plane. As can be seen for these orbital configurations some detectors will reach temperatures in excess of -10°C at some point in the orbit. But even under these extreme conditions, most detectors (75%) will not rise in temperature above -10°C. For all of the orbits and pointing angles between -90-deg and +45-deg the detectors will stay below the maximum allowed temperature [9].

## 3. LAD COLLIMATORS

The purpose of the collimator structure is to provide LAD with its angular sensitivity to celestial point X-ray sources at energies within its operating band and inside its field-of-view, while effectively suppressing all source X-rays incident at larger angles. The collimator is a major component of the mechanical and thermal design, while also acting as a source of particle-induced background. The design is based on a low-mass X-ray collimator, using the technology of microchannel plates (MCPs) and thus drawing on the heritage of the EXOSAT (1983-6) MEDA and GSPC detectors and on the much more recent developments for the BepiColombo Mercury Imaging X-ray Spectrometer (MIXS). In particular, the collimator channel of the MIXS instrument (MIXS-C) provides a secure, high TRL (Technology Readiness Level) basis for the LAD collimator design.

The science requirements which have driven the LAD collimator design are (i) the field-of-view (ii) the required transparency of the collimator at high (i.e. 30-50 keV) X-ray energies and (iii) a possible requirement for a "flat-top" addition to the basic triangular collimator response function, describing X-ray transmission versus off-axis angle. The baseline design has been informed by analytical and Monte Carlo (GEANT4) modelling, by specific X-ray measurements on thick MCPs in the University of Leicester long beamline facility and in discussion with a CP technology provider. The basic collimator units will consist of 8 x 11 cm$^2$ tiles, fabricated from lead glass, with parallel 83 micron square cross-section channels, 5mm long (see a representative example in Figure 10). The open area fraction will be 70%, giving a collimator mass of ~5kg.m$^{-2}$. The details of the baseline design are given in Table 1.

Table 1: CP specification

| PARAMETER | VALUE |
|---|---|
| **Pore size (squared)** | 83 um |
| **Septal thickness (microns)** | 17 |
| **Chanel aspect ratio** | 60: 1 |
| **Channel length** | 5 mm |
| **Open area ratio** | 70% |
| **Channel coating** | None |
| **Leakage** | Acceptable from ~ 30 keV |
| **Pore-to-pore co-alignment** | 1 arcmin |

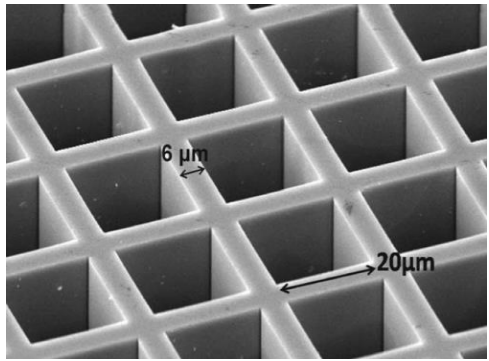

Figure 10: Example of a Collimator design with square pores.

In terms of technical maturity, the baseline collimator technology is already at TRL 4-5 by virtue of the MIXS-related development. We have established that no new CP glass composition is required for LAD, nor slumping of the individual MCPs, nor coating of the channel walls with any metal.

The modular nature of the LAD design also allows the introduction of special collimator units to assist in the separation of the cosmic X-ray background from the particle-induced background in the LAD SDDs. For example, changing the collimator OAR changes the ratio of X-ray to particle events, and allows calibration of deadtime effects. The first "variable transmission" X-ray filter, with a range of OARs in one collimator to assist the dynamic range calibration of LAD, has been fabricated using additive manufacturing techniques.

## 4. LAD ELECTRICAL DESIGN

### 4.1 Overview

The LAD electrical configuration is composed of the following components: the SDD + FEE assemblies, the Module Back End Electronics (MBEEs), the Panel Back End Electronics (PBEEs) and the Instrument Control Unit (ICU) which includes the Data Handling Unit (DHU), the mass memory and the Power Distribution Unit (PDU; see Figure 11).

The electrical architecture of the LAD is controlled by the DHU. The DHU provides an interface to the spacecraft OBDH (On-Board Data Handling) system and also control of the LOFT instrument sub-systems. At the heart of the data handling unit is the Leon processor. The main functions of the DHU are to interface with the PBEEs and the spacecraft OBDH, to configure the instrument and process the engineering and science data (compression and packetisation) and the control of the mass memory. The electrical configuration of the LAD is summarised in Figure 11. A Module Back-End Electronics unit is located inside each detector module and connected to a centrally located PBEE.

Depending on the final configuration of the instrument (with regards to the number of panels and the number of modules per panel) it might be advantageous with respect to the harness to have two or more PBEEs per panel, each located at the center of several modules. The weight of the additional PBEE has to be traded-off against the reduction of harness weight while the power consumption is not significantly different as it scales (per PBEE) roughly proportional to the number of connected MBEEs.

As shown in Figure 11, the PBEEs would be connected to the DHU on a daisy-chained SpaceWire Bus.

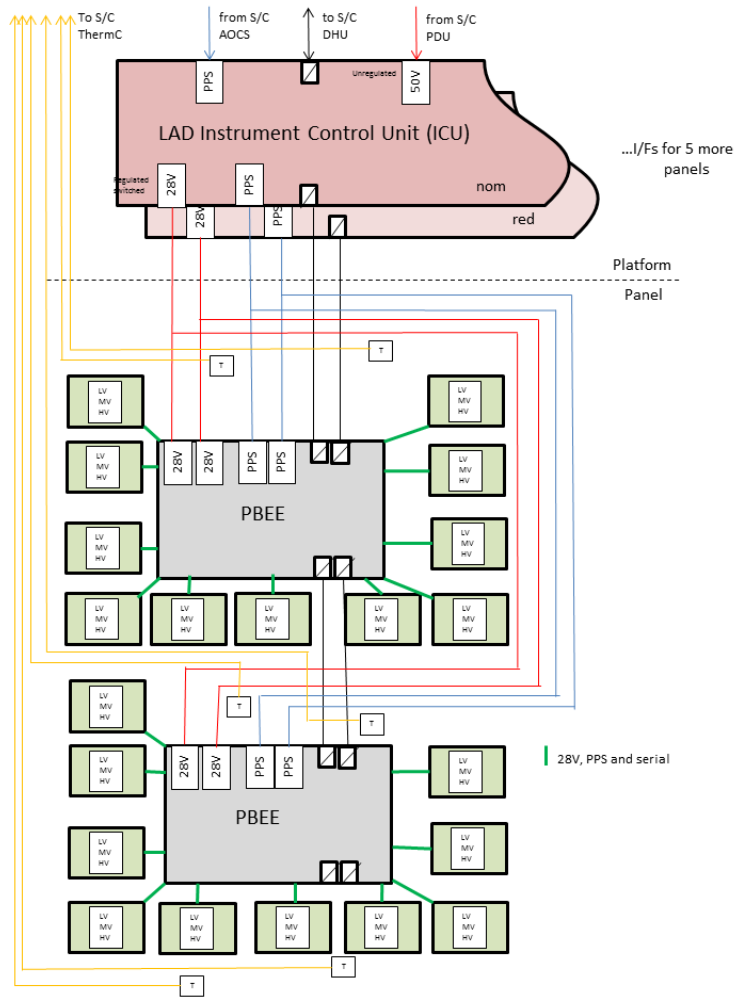

Figure 11: Electrical configuration of the LOFT-LAD.

## 4.2 Silicon Drift Detectors

The LAD detectors are designed on the heritage of the detectors used in the Inner Tracking System of the ALICE experiment at the Large Hadron Collider [3]. The detectors are double-sided, having an area of approximately 100 cm$^2$ (maximum size allowed by a 6-inch wafer). The working principle is summarized in Fig 12: the charge generated by the absorption of an X-ray photon is collected in the middle plane of the detector thickness and then drifted towards the read-out anodes on one edge of the detector. The electric field is sustained by a series of cathodes on both sides of the detector. The LOFT SDD functioning requires a 1300V polarization. Table 2 summarizes the main characteristics of the LAD SDDs.

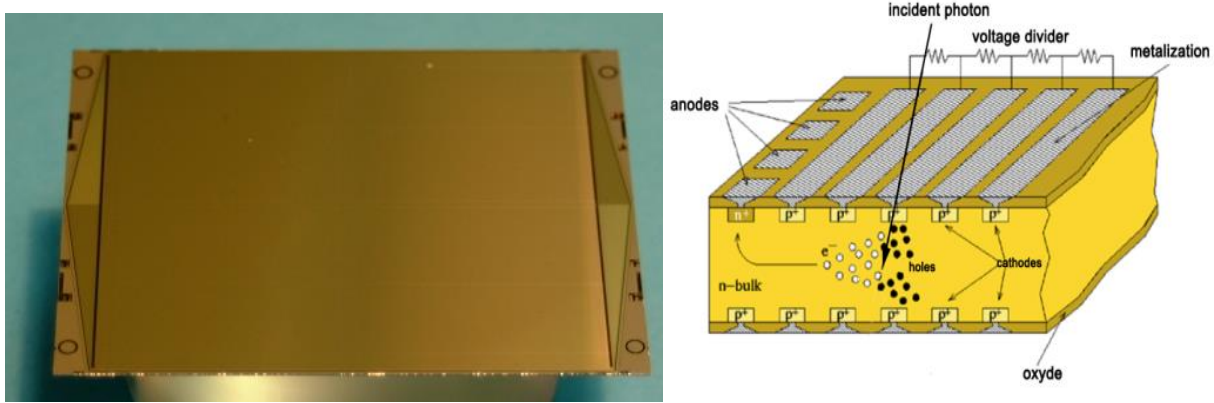

Figure 12: (Right) the ALICE detector (sensitive area 53 cm$^2$); (Left) the working principle of the SDD.

Table 2: LAD SDD main parameters.

| Parameter | LAD |
|---|---|
| Thickness | 450 μm |
| Geometric Size | 120.3 mm x 72.5 mm |
| Active Area | 108.5 mm X 70.0 mm |
| Anode Pitch | 970 μm |
| Drift length | 35 mm |
| Drift field | 360 V/cm |
| Number of anodes | 2 x 112 |
| Anode capacitance | 350 fF |

Based on the ALICE design, a new 76 cm$^2$ sensitive area linear multi-anode silicon drift detector (SDD) [2] has been developed for LOFT [2,5]. The ALICE physics goal required from the SDD design to be optimized for high-resolution, two-dimensional tracking of high-energy ionizing particles in a high-multiplicity environment such as that produced in Pb-Pb collisions at the LHC. The ALICE SDD naturally shares with all other SDDs a feature that makes it very attractive as well for low-energy X-ray spectroscopy applications: its anode capacitance of the order of 100 fF allows for a very low noise contribution from the front-end electronics. On the other hand, when dedicated to low energy X-ray spectroscopy SDDs have a small sensitive area (order of few square centimeters) due to the need to reduce their leakage current, which becomes the primary contributor of noise, and the limiting factor of the energy resolution. The mass production by Canberra Semiconductor, Belgium, of more than 600 detectors in two years permitted to improve the production technology. By the end of the campaign the leakage current at the readout anodes reached values well below 1 nA/cm$^2$ at room temperature for 300 μm thick wafers. All SDDs were tested and the detectors that passed the qualification procedure were assembled into two internal layers of the ALICE ITS [4]. This experience provided a solid basis for the design of a new large sensitive area detector with good spectroscopy performances. During three years of activities we have produced a number of prototypes converging to the final design of the LOFT Large-Area Detector SDD (LAD)[2]. The LAD sensors feature the following characteristics: 2-50 keV energy range with a spectral resolution below 240 eV at 6 keV, power consumption below 0.5 mW/cm$^2$ of sensitive area at room temperature, enhanced quantum efficiency at the low end of the energy range (~39% at 2 keV, maximum of 97% at 9 keV), and the possibility of X-ray spectroscopy timing on a ten microseconds scale.

The final LAD prototype design, delivered by the end of 2013, has been manufactured by Fondazione Bruno Kessler in Trento, Italy, under design by INFN-Trieste. The detector production was carried out using 6-inch diameter floating-zone (FZ) Silicon wafers, a resistivity of approximately 9 kΩ×cm and a thickness of 450 μm. LOFT is required to cover large sensitive areas, even not continuous, and the sensor geometry is not a critical issue. For this reason the ALICE SDD design was kept as a basis for the new detector geometry. It enables an easy tiling of single SDDs into a large area. The silicon tile containing the SDD is a rectangle with an area of approximately $120 \times 72$ mm$^2$, i.e. the largest cut out from a 6-inch diameter wafer without changing the drift length. Even though the cut out is a rectangle, the detector itself has a hexagonal shape with a rectangular sensitive area of about $108 \times 70$ mm$^2$. The detector is symmetrical with respect to the central p$^+$ cathode, i.e. it has a bi-directional structure, where electrons drift from the central p$^+$ cathode towards two linear arrays of readout n$^+$ anodes. For each half-detector there are 292 drift cathodes with a pitch of 120 μm and 112 readout anodes with a pitch of 970 μm. The sensitive-to-total-area ratio is 87 %. Aside from the integrated dividers there are two triangular areas constituted by the guard p$^+$ cathodes. There is one guard cathode every two drift cathodes, so the potential difference between adjacent guards is twice that between adjacent drift cathodes. The pitch of the guard cathodes is 32 μm. The collection zone of the detector consists of only 3 cathodes biased independently from the integrated voltage divider: the 'grid' cathode on the n-side separates anodes one from another, while two 'kick up' cathodes on the opposite p-side force the drifting charge towards the anodes array. The detector is planned to work at a drift field of 360 V/cm (HV bias of -1300 V), entailing a maximum drift time of about 7 μs at 20 °C, which is reduced to ~5 μs at -20 °C due to the higher electron mobility at lower temperatures.

### 4.3 FEE electrical design

The LAD FEE electronic architecture is presented in Figure 13.

The FEE is separated in 2 symmetrical rows of 7 ASICs (see [8] for all details on the ASIC design). Each ASIC has 16 inputs connected to the SDD anodes. Each ASIC of the row is chained by a differential trigger line allowing trigger to be propagated from the 1$^{st}$ ASIC to the last ASIC and finally to the MBEE. The MBEE is also connected to the trigger input of the 1$^{st}$ ASIC and can force a trigger to all ASICs in order to process some noise measurements. Each ASIC also has an individual hardwired address allowing its individual access through the common command (CMD) and clock (CLK) differential signals where individual ASIC address (0 to 6) or broadcast (7 i.e. all ASICs) can be set within the command frame. Reset and Hold signals are also common to each row. Data outputs are divided into 2 differential pairs of signal per FEE side: Odd and Even DOUT. This separation allows a higher bandwidth when all ASICs send their data especially when a trigger occurs on channels shared by two adjacent ASICs: in this case, a single DOUT frame can be processed simultaneously on the 2 ASICs.

The 2 ASIC rows are connected to 2 Hyperstac connectors located at the end of the flex extensions: one 40 pin connector for low voltage digital and power supply signals and another 4 pin connector for the SDD high voltage and medium voltage signals. Hyperstac connectors have been chosen for their simplicity of connection and cost: no soldering operation is required during manufacturing. High voltage can also be passed through them thanks to a specific design for LOFT performed by the manufacturer with appropriated insulation distances. Finally, a thermocouple is mounted in the centre of the board for housekeeping measurements.

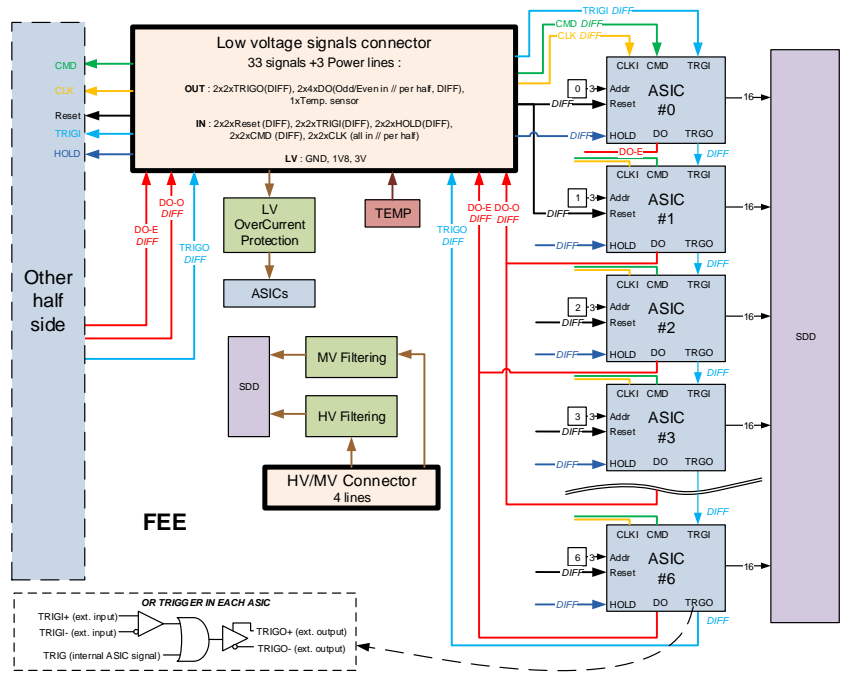

Figure 13: LAD FEE Electronic architecture

## 4.4 MBEE and PBEE electrical design

When the collected charge exceeds a programmable threshold in one of the ASIC channels in the Front-End Electronics (FEE), a trigger signal is forwarded to the Module Back-End Electronics (MBEE) where a time tag (based on a precise 1 MHz clock count) is instantaneously generated. The trigger is also propagated directly from one ASIC to all of the ASICs on the respective detector half and they all freeze their current signal. The MBEE requests the trigger map from all of the ASICs on this detector half and validates if only one or two adjacent anodes triggered. If the event passes this selection criteria, the "A/D conversion" command will be sent from the MBEE to the triggering ASIC and the conversion will be carried out (inside the ASIC), providing an 11-bit output per anode.
Following the A/D conversion, the MBEE processing pipeline will be activated. The saved time tag will be added to the event package at the end of the processing pipeline. The main processing functions of the MBEEs are:

- time tagging
- trigger validation and filtering
- pedestal subtraction
- common noise subtraction
- energy reconstruction
- event threshold application
- Housekeeping data collection.

Each MBEE consists of two PCBs and is designed to process events within a pipeline structure that handles the events from one side of a detector (7 ASICs). This pipeline is initiated 32 times within the MBEE FPGAs to allow the processing of data from all 16 detectors simultaneously. The pipeline is designed such that the processing time within each step is the same and shorter than the A/D conversion time of a following event. In this way, the data processing in the MBEE does not inflict additional dead-time and the pipeline is always ready for the next event.

Due to the large number of MBEEs on the LAD a very lightweight whilst robust connection to the Panel Back-End Electronics (PBEE) is required. The communication interface is designed as a bi-directional point-to-point connection between the PBEE and each MBEE. It is using the SpaceWire hardware standard but does implement a custom, SPI-like data transfer running at 10 MHz.

During the study phase, a functional PCB Prototype of the MBEE was designed and manufactured, allowing communication to four Silicon Drift Detectors (SDDs). An additional FPGA is part of the PCB board, allowing the simulation of up to four additional SDDs. This PCB can be used to verify the communication between the MBEE and the PBEE, and test the different state transitions within the MBEE, as well as the complete processing pipeline VHDL code. The now fully functional pipeline was tested with simulated event data to verify the individual steps in the data processing. With the current design, the MBEE can handle up to 8 parallel processing pipelines.

The PCB design considers the designated Microsemi RTAX 2000 FPGA footprint on the PCB board, although this FPGA is not suitable as a developing platform, as it can only be written once. A solution for this development stage is the Aldec Adapter, using a ProASIC reconfigurable FPGA.

The Panel Back-End Electronics (PBEE) handles all events from the individual modules of one of the six detector panels (depending on the design up to 25 modules). It is the heart of the data acquisition and signal processing chain, located between the individual modules and the DHU on the satellite bus. The main tasks of the PBEE are:

• Interfacing the MBEEs
• Collecting and buffering the event packets
• Differential time assignment
• Reformatting the data to binned data depending on the observation mode
• Transferring the data to the DHU
• Collection of HK data and creation of HK packets.

Power will be distributed in the form of 28 Volts from the S/C PDU via the DHU and the PBEEs to the individual MBEEs. All necessary lower voltages are generated at the respective level (DHU, PBEE, MBEE) by local power converter boards. This design allows for a minimum of harnessing over the panel hinge (where the bending radius is important) and from the PBEE to the MBEEs. The high voltage necessary to operate the detectors is also generated by a dedicated HV-daughterboard in the individual modules.

There is no redundancy at the PBEE level as the loss of one of the panels would not compromise the scientific goals of this mission. In the case of a SC design with fewer panels, segmentation of the PBEE is necessary in order to maintain an appropriate level of graceful degradation in the event of a single PBEE failure. The interface between PBEE and DHU is a fully compliant SpaceWire interface, running at 100 MHz.

The IAAT has developed, designed and built a PCB Prototype of the PBEE that is capable to receive events from up to 10 individual MBEEs and additionally incorporates an MBEE simulator to simulate another two MBEEs. The board utilises a Xilinx Virtex 4 FPGA running at 100 MHz. The PBEE and the MBEEs create an additional 10 MHz clock which is then used for the interface. The connectors are currently 9 pole Sub-D connectors, but will be replaced by SpaceWire connectors in the final design. The on-board simulator consists of a Xilinx Spartan 3 FPGA. The board is equipped with a SpaceWire port, enabling the PBEE to send the processed event data via space-wire to a lab PC. The PC can send commands to the PBEE and configuration data to the MBEEs.

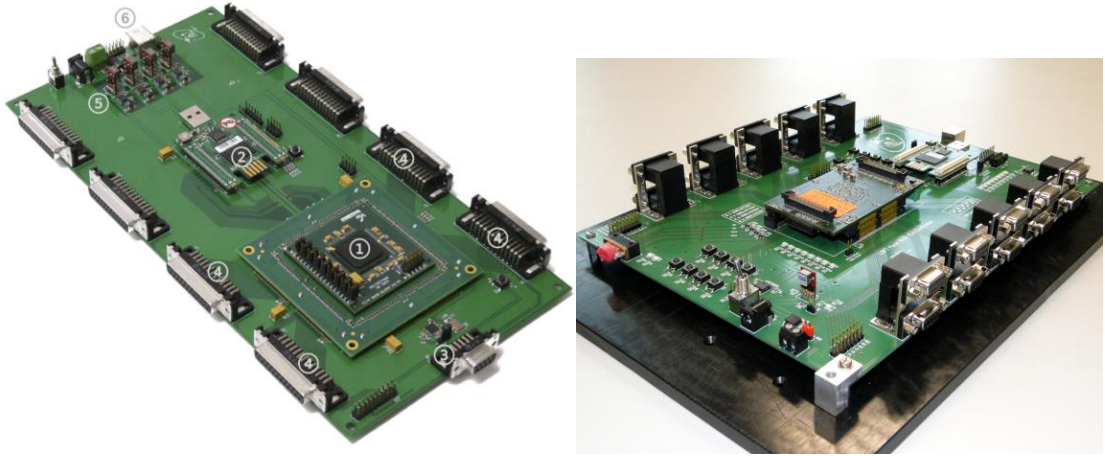

Figure 14: Left: Prototype of the Module Back-End Electronics hosting the central processing FPGA (1), an ASIC simulator (2), a connector to the PBEE (3), connectors to the Front-End Electronics (4) and LV Power Supplies (5). Right: Prototype Board for the Panel Back-End Electronics including programmable boards (a Virtex-4 FPGA for central data processing and a Spartan-3 FPGA for MBEE simulation) and connectors to 20 MBEEs (left and right) as well as to the ICU (left side, red cap).

### 4.5 Instrument Control Unit

The Instrument Control Unit (ICU) forms the major controlling element of the LAD instrument. It provides the interface to the spacecraft OBDH and also control of all LOFT instrument sub-systems via SpaceWire. The ICU box consists of three components: the Data Handling Unit (DHU), the mass memory and the Power Distribution Unit (PDU). The tasks performed at the level of the ICU involve telecommand execution and relay, access to mass memory, time distribution, data processing and compression, HK collection, instrument health monitoring and calibration tasks.

The ICU box is provided twice for cold redundancy. At its heart is the Data Handling Unit where the scientific data is compiled, processed and compressed. A LEON III processor was chosen for the task to run the onboard software (see next section) due to its additional flexibility when compared to a hardware only architecture, i.e. a state-machine. The DHU collects the data from the PBEEs via a dedicated SpaceWire interface PCB, performs the selection and formatting tasks depending on the selected observation mode and commits the data to the mass memory in a compressed format. Besides the routing and execution of telecommands, the ICU can run macro commands for several tasks, including on-board calibration and the individual sequenced switching of the MBEEs in the engineering modes.

The time distribution for the LAD will be based on a highly accurate 1 MHz clock provided by the DHU, routed down to the PBEEs and MBEEs and synchronized once per second by the DHU with the Pulse Per Second (PPS) received from the GPS.

## 5. SOFTWARE

The LAD on-board software runs on a LEON-3 processor and its main functions are instrument control and monitoring and science data processing and formatting. Software is designed to allow the instrument to have the complex functionality that is necessary to allow itself to be updated and work around problems automatically and with input from the ground. Instrument control will be possible through the software via telecommands from the ground (e.g., power on & off, set-up of ASICs and FPGAs, loading parameters for processing/on-board calibration, investigations) and autonomously on-board (e.g., mode switching, FDIR and diagnostic data collection). The software implements the standard ECSS-style PUS service telecommand packets for housekeeping, memory maintenance, monitoring, etc., and some standard telemetry packets for command acceptance, housekeeping, event reporting, memory management, time management, science data etc.

The software needs to be able to send set-up information to the hierarchy of processing elements and receive and process the housekeeping data coming back, simplifying these data in a configurable way for a lower rate transmission to the ground.

The software will be written in C using RTEMS, the real-time executive, which will schedule the software tasks, each at different priorities, communicating with each other with messages and events. The software package will consist of two separate sub-systems. "Basic" software, stored in a very reliable PROM(s), would have enough functionality to perform basic health management and memory management: to receive, store and execute patches or new software. "Operational" software, stored in EEPROM, would have the functionality of the "Basic" software and also the full science capabilities. In this way new software can be loaded to the instrument without overwriting or corrupting the original PROM code. If there is any problem with the software interface to the spacecraft, a reboot or power off/on of the instrument will reset the software into the well-tested "Basic" mode which does not produce science data, thereby initialising the instrument back to a well-defined mode. As the software has to operate in a remote space environment, it will be written to be robust against errors, to report as much information as possible on any problems encountered and progress made (to help investigations) and perform any operations required by EDAC/memory scrubbing. Error messages will be carefully controlled so that the Spacecraft is not flooded with messages that may have a common cause. A software-controlled hardware watchdog will be implemented.

The interfaces between the software and the rest of the instrument/satellite will be as clean as possible with the processor and software taking over the processing of the events and diagnostic data at the point where the data streams from the panels are joined. Interaction with the Spacecraft will be through SpaceWire (with the option of CAN).

Time information will be received from the GPS and distributed via the PBEEs to the MBEEs. The MBEEs will be monitored in case re-synchronisation is needed.

The DHU will collect data from housekeeping sensors in order to monitor configuration status and instrument health. Housekeeping will be polled on a regular basis; frequency will be dependent on exact function.

Commandable modes include:

- Science: Including user-defined energy and time, action to be taken during SAA passage and earth occultation
- Engineering: Including on-the-fly-configure, diagnostic-and-calibration, pedestal, electrical-calibration
- Non-operating modes: Off, Safe

These modes and other operations will be implemented in accordance with ECSS-E-70-41A - ECSS service numbers in square brackets:

- Housekeeping control [3] (rate and possibly content)
- Memory management [6] (load, dump, checksum of absolute address)
- Function management [8] (high voltage control, parameter loading, mode changes)
- Time management [9] (request time from instrument software, resynchronise time)
- Monitoring [12] (limit checking, reporting and actions)

In addition, the on-board software is able to make autonomous decisions, for example to go to Safe mode, switch off out-of-limits detectors or modules, organise data for telemetry.

*Software science functionality*: The software will collect and format the acquired science event data. At the lowest level it will be able to monitor and set-up the registers of each of the electronic elements of the event processing hierarchy. The data from the 6 panels will arrive at the processor board and be handled by dedicated electronics under the control of the software. The resulting data stream will have the remaining processing done by binned as necessary, reformatted ready for compression and packetised. The software will interact with the Spacecraft, sending the data over SpaceWire for transmission to the ground. The functionality includes:

- Set-up and monitoring of detectors disabling portions of the electronics if necessary
- Pedestal measurement
    - Force triggers
    - Collect data
    - Calculate values
- Electrical calibration
    - Gain & off-set
- Manage the data compression and packetisation

- Manage the data prioritisation and storage including if necessary for re-transmission
- Time management
    - Set the time on the MBEEs as they are powered
    - Check the MBEE synchronisation every second
    - Resynchronisation if necessary
- Observation management: autonomously observing following a plan

*Software data storage*: The following data storage requirements are foreseen:

- 450 KByte required for detector definition
    - This includes 128 bit per ASIC: fine threshold, enable channels. It should be possible for this to be telecommanded to the instrument.
- 2.7 MByte for calibration data
    - This is 1 pedestal and RMS for each channel and gain per channel. It should be possible for this to be telecommanded to the instrument.
- 210 GByte for science storage.
    - I.e. enough for more than 1 orbit (as ground-passes can be lost) and enough to cope with a high count rate for a short time

The LAD modes diagram is illustrated in Figure 15.

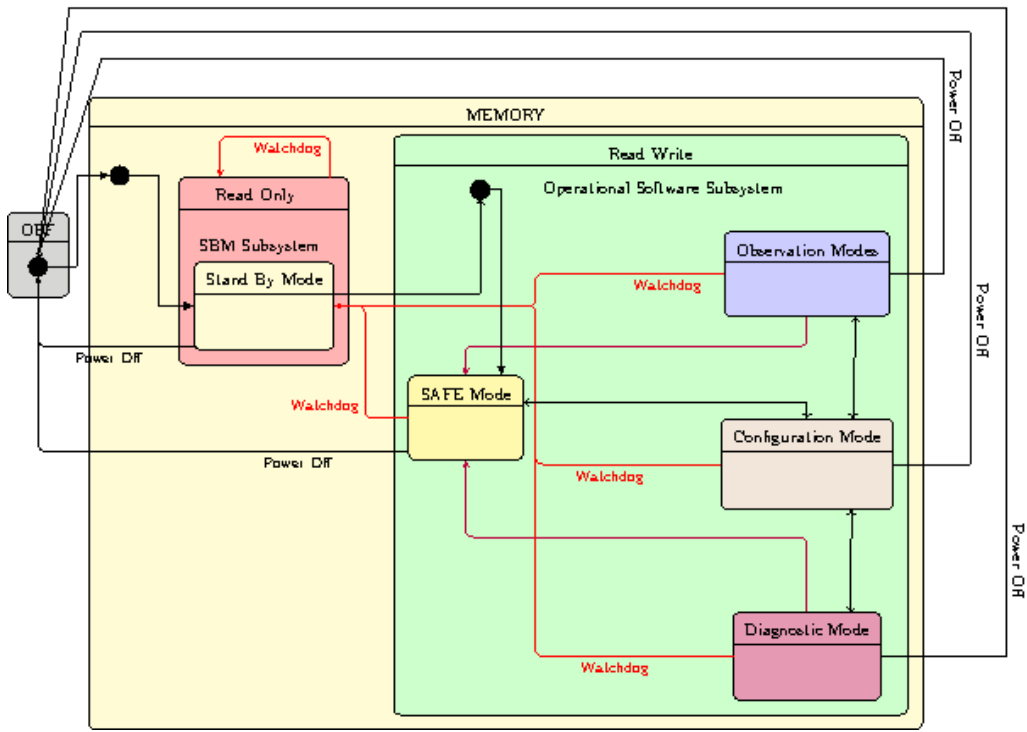

Figure 15: Schematic view of the LAD modes diagram.

# 6. CALIBRATIONS

The calibration plan of the LAD foresees two types of activities:

- On-ground calibration
- In-orbit calibration

where the plan is to calibrate as fully as possible on the ground in order to minimise the loss of observing time in orbit.

There are five key aspects of the LAD performance, detailed in the science requirements table, which require calibration, and this is summarized in Table 2.

Table 2: Calibrations required

| Parameter | Specification |
| --- | --- |
| Deadtime as f(countrate) | <1% @ 1 Crab |
| Knowledge of energy scale | $10^{-2}$ |
| Energy resolution as f(Energy) | FWHM @ 6 keV: Singles 200 eV, overall 240 eV |
| Collimator performance as f(Angle, Energy) | Table 1, and background < 10 mCrab |
| Effective area as f(Energy) | 10 $m^2$ @ 8 keV, known to 15% |

## 6.1 On-ground Calibration

The parameters in Table 2 can all be calibrated in ground measurements, and fall into two categories depending on the fact that they require either uncollimated or collimated calibrations. In particular, deadtime, knowledge of energy scale and energy resolution do not rely on collimator performance and therefore are most efficiently performed at detector tray level (where the countrate is not reduced by the presence of the collimator). The performance can be measured using radioactive and fluorescence sources and, during these tests, the ASIC stim pulses can also be calibrated.

Collimated calibrations are instead required for the last two parameters in Table 2. Collimator performance can be measured using a combination of metrology (physical and optical) and X-ray measurements, at tile and collimator tray level. The effective area is a combination of collimator throughput (intrinsic pore collimation and alignment effects at various levels: pore-to-pore, tile-to-tile, module-to-module and panel-to-panel) and detector QE. The effective area calibration will be therefore performed at module level in a consortium beamline, combined with optical metrology of fiducial reflectors on the collimator frame. Module-to-module and panel-to-panel alignment effects will be measured using optical metrology methods at the industrial facilities used for panel and s/c level integration.

AIV considerations:

The scale of the LAD instrument means that the entire instrument production and test programme has to be pipelined. The proposed plan was to perform calibration activities at module-level (or small groups of modules), in sequence with manufacture and delivery. As with other aspects of the instrument production, performing calibration on modules makes the activity feasible in reasonable-sized facilities. Figure 16 is a section of the Work Breakdown Structure proposed by the LAD consortium, showing how calibration activities are interleaved with AIV activities. Numbers alongside the boxes show the number of times each activity has to be performed.

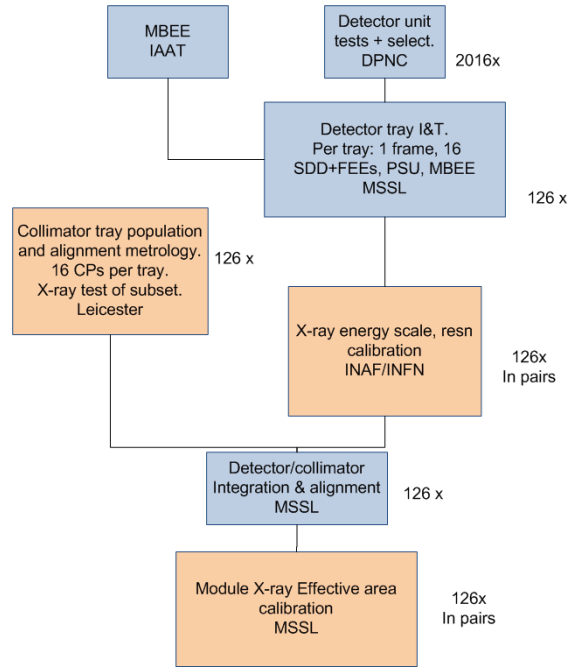

Figure 16: Portion of Work Breakdown Structure proposed by the LAD consortium, showing interleaving of AIV and calibration activities

## 6.2 In-flight calibration

5% of instrument lifetime has been identified for calibration activities (with a goal of 2%). In-flight calibration activities consist of the following:

- Repeated observation of known sources
- Measurement of parameters for processing (using stim pulses and threshold scans). These include pedestal (download of 1.9 Mbyte) and electrical calibration data (download of 1 Gbyte).

In addition to specific in-flight calibration activities, simulations show that fluorescence of Pb L lines (10 keV and 12 keV) from the collimator glass and rear shield will produce detectable lines which can be compared with spectra.

## 7. FUTURE DIRECTIONS

As previously mentioned, the LOFT mission has not been down-selected for launch within the M3 call. However, a long (>2 years) assessment study, has culminated with the delivery of a Yellow Book and of more than 3000 pages of technical documentation. Most of the trade-offs have been closed, leading to a robust design.

Based on the ESA technical and programmatic review, on the scientific reports by the Advisory Structure and on the Q&A session with the Astronomy Working Group and Fundamental Physics Advisory Group, the feedbacks for LOFT in M3 have all been very positive. In particular, the LOFT science was fully recognized as very strong, the mission suitable to address it and the technology was evaluated as mature. The non-overlap between the LOFT science and the science case of the Athena mission was also clearly acknowledged.

The high level of readiness and maturity of the mission and payload design, as well as the clean and solid assessment of the unique science case still make LOFT a very competitive mission with a compelling science case, and for this reason the LOFT Consortium Board has recently confirmed the intention to continue the development aiming now at the upcoming M4 ESA launch timescale (2026).

The work of the MSSL-UCL and Leicester groups is supported by the UK Space Agency. The work of SRON is funded by the Dutch national science foundation (NWO). The work of the group at the University of Geneva is supported by the Swiss Space Office. The Italian team is grateful for support by ASI (under contract I/021/12/0), INAF and INFN. The work of IAAT on LOFT is supported by Germany's national research center for aeronautics and space - DRL. The work of the IRAP group is supported by the French Space Agency. LOFT work at ECAP is supported by DLR under grant number 50 00 1111. RH acknowledges GA CR grant 13-33324S and MSMT LH13065.